\documentclass[fleqn,10pt]{wlscirep}
\usepackage[utf8]{inputenc}
\usepackage[T1]{fontenc}
	\usepackage{makecell}
\title{Projecting XRP  price burst by correlation tensor spectra of transaction networks}
\author[1,2,*]{Abhijit Chakraborty}
\author[2,+]{Tetsuo Hatsuda}
\author[1,$\dag$]{Yuichi Ikeda}

\affil[1]{ Kyoto University, Graduate School of Advanced Integrated Studies in Human Survivability, Kyoto, 606-8306, Japan}
\affil[2]{RIKEN Interdisciplinary Theoretical and Mathematical Sciences Program, Saitama, 351-0198, Japan}

\affil[*]{chakraborty.abhijit.7y@kyoto-u.ac.jp}
\affil[+]{thatsuda@riken.jp}
\affil[$\dag$]{ikeda.yuichi.2w@kyoto-u.ac.jp}

\begin{abstract}
Cryptoassets are becoming essential in the digital economy era. XRP is one of the large market cap cryptoassets. Here, we develop a novel method of correlation tensor spectra for the dynamical XRP networks, which can provide an early indication for XRP price. 
A weighed directed weekly transaction network among XRP wallets is constructed by aggregating all transactions for a week. A vector for each node is then obtained by embedding the weekly network in continuous vector space. From a set of weekly snapshots of node vectors, we construct a correlation tensor. A double singular value decomposition of the correlation tensors gives its singular values. The significance of the singular values is shown by comparing with its randomize counterpart. The evolution of singular values shows a distinctive behavior. The largest singular value shows a significant negative correlation with XRP/USD price. We observe the minimum of the largest singular values at the XRP/USD price peak during the first week of January 2018. The minimum of the largest singular value during January 2018 is explained by decomposing the correlation tensor in the signal and noise components and also by evolution of community structure. 
\end{abstract}
\begin{document}

\flushbottom
\maketitle
%\thispagestyle{empty}
%\noindent Please note: Abbreviations should be introduced at the first mention in the main text – no abbreviations lists. Suggested structure of main text (not enforced) is provided below.
\section*{Introduction}
%%%%%%%Crypto assets and its relavance 
Cryptoassets represent value that one can transfer, store, or trade digitally. It uses cryptography to protect data and distributed ledger technology to record transactions. A blockchain technology, which is a form of secure digital ledger, is used to store the records of crypto transactions. 
Recently, cryptoassets have been very popular as an investment, but cryptoasset prices are extremely volatile and unpredictable. The cryptomarket experienced severe price fluctuations from 2017, December to 2018, January.   The presence of bubbles i.e., explosive price behavior in this asset has attracted attention from the researchers. 
At present, there are many cryptoassets. Some well-known of them are Bitcoin (BTC), Ethereum (ETH) and XRP. For the last one decade complex network theory has been widely used to analyze cryptoasset transaction data. Among the cryptoassets, networks of BTC and ETH transactions have been studied extensively~\cite{wu2021analysis}, which include different structural properties~\cite{kondor2014rich, ferretti2020ethereum}, temporal evolution~\cite{kondor2014rich, tasca2018evolution, liang2018evolutionary, 9142770} and market effect~\cite{kristoufek2013bitcoin, koutmos2018bitcoin,ciaian2016economics} of these networks.   
In contrast, XRP transactions have been less explored.
Ripple Labs Inc. created XRP as the native cryptocurrency for the Ripple network, which is designed to provide fast, efficient, and cost-effective financial transactions. The Ripple network is used for real-time gross settlement of financial transactions, currency exchange, and cross-border remittances. The goal of the Ripple network is to provide a stable and decentralized ledger system that can be used to facilitate cross-border transactions in a more efficient and cost-effective manner. XRP is used as a bridge currency in the network, helping to facilitate cross-border transfers and providing liquidity to the system.
%Ripple Labs Inc., a US-based technology company, created XRP as a native currency for the Ripple network, a real-time gross settlement system, currency exchange and remittance network with aims of fast transactions settle, low-cost transaction fees, scalable, stable, sustainable and decentralized ledger system. 
P. Moreno-Sanchez {\em et al.} uncovered community formation and clustering properties for the Ripple network~\cite{moreno2018mind}. Y. Ikeda studied the structural properties of XRP transaction network, which include the heavy tail nature of nodal strength distributions, low value of clustering coefficient and significant triangular motif~\cite{ikeda2022characterization}. More recently, the role of the most active nodes with respect to outgoing and incoming flows for a duration including bubble/crash period has been quantified for BTC and XRP transaction networks~\cite{aoyama2022cryptoasset}.

%What is correlation and how useful is it? 
Any statistical dependence between a pair of variables
can be studied by cross correlation which is predictive and useful for empirical data. 
It can be measured in different ways: One of the simple and well-known methods is the Pearson correlation which represents the linear dependence between variables and is defined for a pair of variables $x$ and $y$ with $n$ observations as $\rho=\sum\limits_{i=1}^n (x_i-\overline{x})(y_i-\overline{y})/((n-1)\sigma_x \sigma_y)$. Here $\overline{x}, \overline{y}$, and $\sigma_x, \sigma_y$ indicate mean and standard deviations of $x$ and $y$. 
%This cross correlation is used extensively for studies of financial time series data~\cite{laloux1999noise}, medical time series data such as electroencephalogram, magnetoencephalography  data recordings~\cite{schindler2007assessing} and for other time series data. The application of cross correlation on stock market data shows the relationship between stock price changes and liquidity or trading volume. 
The cross correlation methodology armed with random matrix theory (RMT) is mostly applied to time series data. The aim of such method is to analyze high dimensional data to find key factors for the collective dynamics of many quantities.  For example, it is used to study the daily returns of different stocks~\cite{laloux1999noise, plerou1999universal, plerou2002random} and foreign exchange rates~\cite{chakraborty2018deviations, chakraborty2020uncovering }, monthly macroeconomic data~\cite{kichikawa2020interindustry} , or different medical data such as electroencephalogram, magnetoencephalography  data recording~\cite{schindler2007assessing}.
Also, D. Kondor {\em et al.} applied principal component analysis on the matrices obtained from the daily network snapshots to show the relationship between the price of bitcoin with structural changes in the transaction network~\cite{kondor2014inferring}. 
The application of cross correlation on stock market data shows the relationship between stock price changes and liquidity or trading volume~\cite{podobnik2009cross}.
Note that these are time series of  variables that emerge due to interactions of different
entities of the system. In most cases, we lack detail information of the interactions at the micro level. For example, we do not have detail information about the interaction among the individuals in stock markets.  
%we introduce we have extended the RMT 
%a new approach for XRP transaction network.
\begin{figure}
\centering
\includegraphics[width=0.75\textwidth]{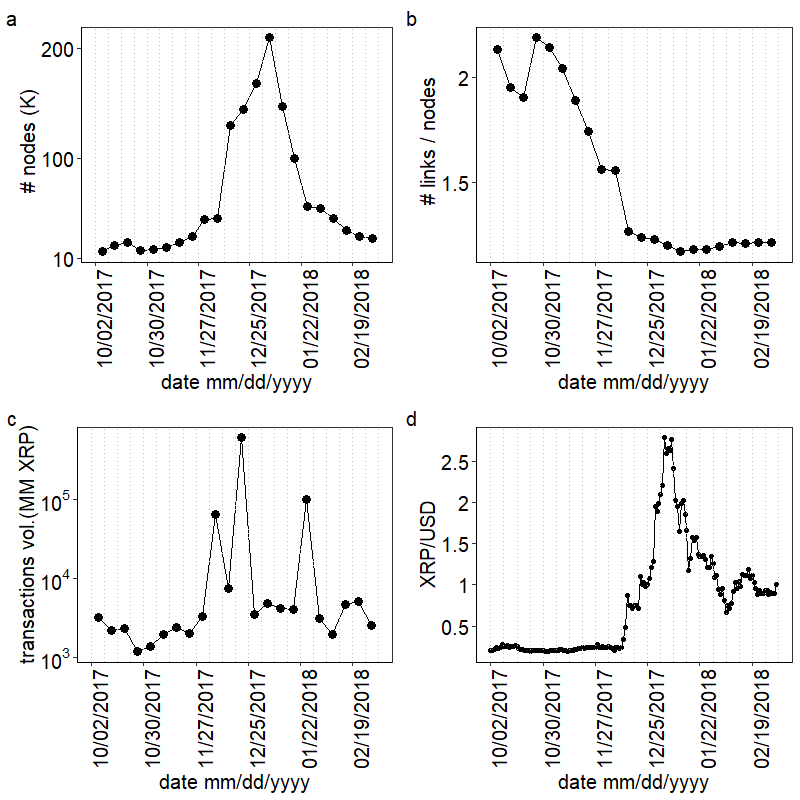}
\caption{ The variation of (a) total number of nodes (b) number of links per node and (c) total transaction volume in millions XRP for each weekly network. (d) The daily XRP/USD close price (source: \url{https://www.marketwatch.com/investing/cryptocurrency/xrpusd}). 
The dotted grey vertical lines represent the weekly windows.}
\label{Fig1}       % Give a unique label
%\vspace{-0.4cm}
\end{figure}

 In the case of XRP, we know the information for all transactions between wallets. Using this high-quality micro level data, we develop a method of cross correlation tensor which can be applied on dynamical XRP transaction networks. Our method relates the structural properties
of XRP networks to the XRP/USD price.
To calculate the correlation tensor, nodes of the networks are converted to vectors using network embedding techniques~\cite{perozzi2014deepwalk, grover2016node2vec}. 
%The network embedding algorithm learns the social representations for the nodes of a network.
We use the DeepWalk embedding technique~\cite{perozzi2014deepwalk}, which uses a set of truncated random walks to learn latent features. The latent features capture neighborhood information and community memberships of the nodes. Using a network as the input, the algorithm provides a latent representation in a continuous vector space as an output. A generalization of the DeepWalk algorithm is node2vec ~\cite{grover2016node2vec}, which uses biased random walks to learn features that encode more complex relations of nodes, such as functional relationships.  Using the vector representation for a subset of nodes present in every weekly XRP transaction networks, we 
calculate correlation tensors for different time periods. We perform a double singular value decomposition on the correlation tensors to obtain its spectra. We compare the results with reference correlation tensors to understand  the significance. As a reference correlation tensor, one usually uses the RMT method. However, as we are using a small time window to calculate the correlation tensor, RMT method will not be suitable as reference. We use randomized and reshuffled correlation tensors as reference correlation tensors. The largest few singular values capture the impact of the  bubble period or crash in XRP price. The largest singular value is found to be significant and has a strong negative correlation with XRP/USD price. Furthermore, it provides an early indication for XRP/USD price, including bubble periods or crashes. 

% on the network structure.  

%Here, we develop a new method of correlation tensor spectra from networks of direct XRP transaction to detect bubbles in XRP price. 
%    
%
%Crypto assets are getting a lot of attentions from investors, regulators and policymakers in recent times because of their large market capitalization and volatile nature in price. XRP is one of them developed by Ripple Labs. Inc.. This crypto asset has experienced severe price fluctuations around the end of $2017$. The presence of bubbles i.e. explosive price behaviour in this asset has attracted attention from the researchers. 
%Here, we develop a new method of correlation tensor spectra from networks of direct XRP transaction to detect bubbles in XRP price. 
%

\section*{Results}
\begin{figure}
\centering
\includegraphics[width=0.48\textwidth]{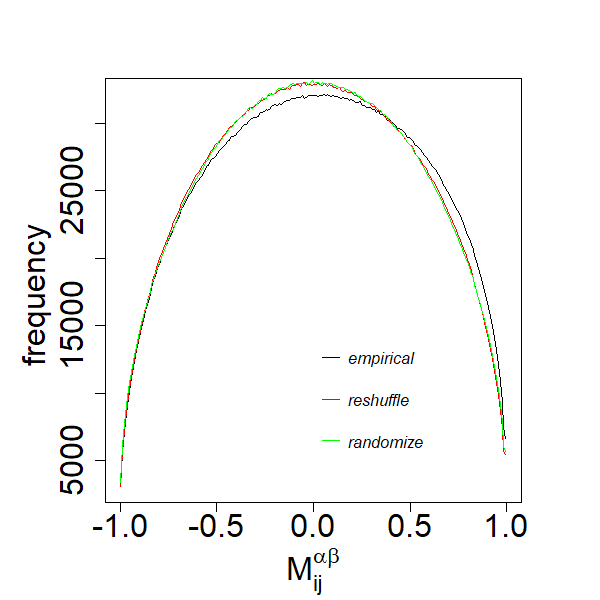}
\caption{The distributions of the elements of empirical, reshuffle and randomize correlation tensors calculated for the week, November 06 to November 12, 2017. The data are averaged over 20 uncorrelated embedding of the networks.
}
\label{Fig2}       % Give a unique label
%\vspace{-0.4cm}
\end{figure}
%%%%%----Fig1
The weekly network obtained from XRP transactions between wallets evolves with time. We focus our study on the duration 2017 October 02 to 2018 March for which covers a bubble period in XRP price. It indicates $22$ weekly networks. The number of nodes for each weekly network is shown in Fig.~\ref{Fig1}~(a). We observe that the number of nodes of the weekly networks increases rapidly from $45, 169$ during the week of 2017, November 27 - December 04 to the peak value $209, 143 $ during the week of 2018, January 01 - January 07 and later further falls to $27, 811$   during the week of 2018, February 26 to March 04. Fig.~\ref{Fig1}~(b) shows the decline in the number of links per node indicating the average nodal in- or out-degree from $2.15$ to $1.20$ during 2017, October 02 - 2018 March 04. The decline in the number of links per node indicates a reduction in the average frequency of transactions of a node with other distinct nodes. We note that XRP transaction volume was very high for three weeks between December 05 - December 24 and also for the week during 2018, January 22 - January 29, as shown in Fig.~\ref{Fig1}~(c) . The three weeks of extremely high XRP transaction volume may have contributed to the bubble formation in XRP/USD during 2017, December 25 to 2018 January 07. The daily XRP/USD price from October 2, 2017 to March 4, 2018 is shown in Fig.~\ref{Fig1}~(d).  The XRP/USD price had an extraordinary  rise and fall between December 2017 and January  2018. This indicates a bubble period for XRP. We consider this period for our study because this is the most significant bubble period for the cryptoasset market. The chart of XRP/USD prices for a more extended period is shown in SI Fig. S1.   
%%%%
\begin{figure}
\centering
\includegraphics[width=0.48\textwidth]{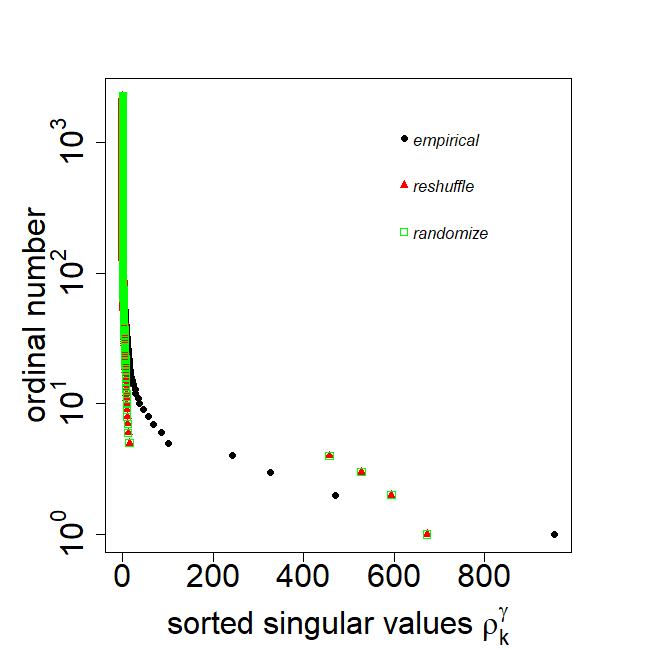}
\caption{Sorted singular values $\rho_k^\gamma$ of the empirical, reshuffle and randomize correlation tensor for the week, November 06 to November 12, 2017. The values represent average over $20$ uncorrelated embeddings of the network.
}
\label{Fig3}       % Give a unique label
%\vspace{-0.4cm}
\end{figure}

\begin{figure*}
\centering
\includegraphics[width=0.98\textwidth]{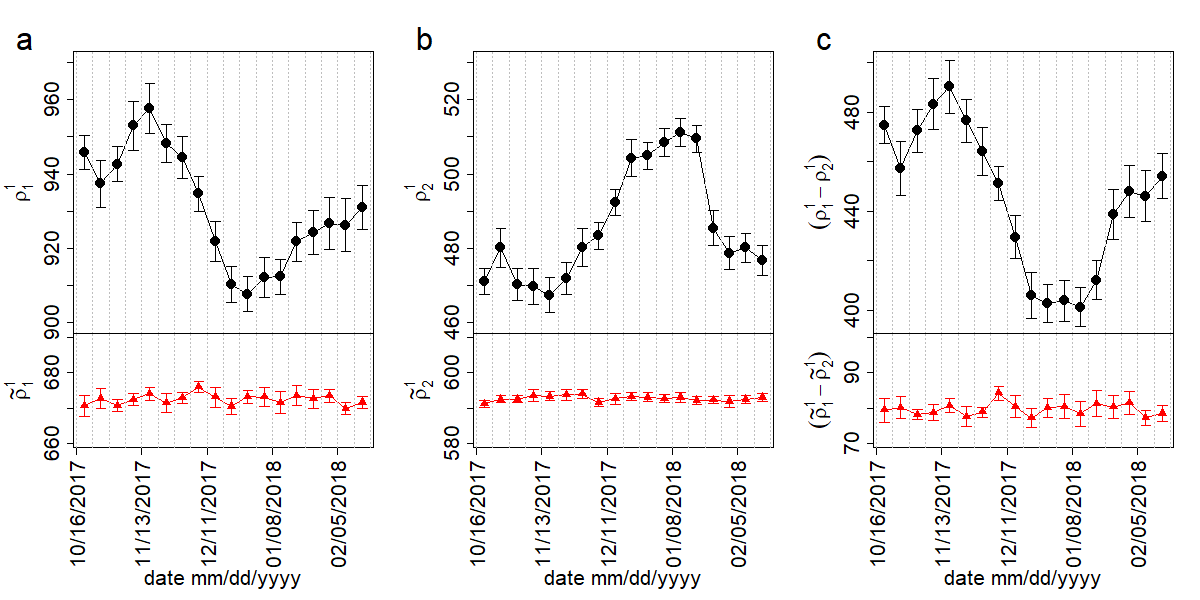}
\caption{ Evolution of the singular values for empirical $\rho_k^\gamma$ and reshuffled $\tilde{\rho}_k^\gamma$ correlation tensors for different weeks. (a) Variation of  the largest singular value for empirical $\rho_1^1$ and reshuffled $\tilde{\rho}_1^1$ correlation tensors.   (b) Variation of the second largest singular value for empirical $\rho_2^1$, reshuffled $\tilde{\rho}_2^1$ correlation tensors. (c) Variation of spectral gap for empirical  $(\rho_1^1 - \rho_2^1)$ and reshuffled  $(\tilde{\rho}_1^1 - \tilde{\rho}_2^1)$ correlation tensors. The error bars indicate the standard deviation. The data are averaged over $20$ uncorrelated embeddings of the networks. The dotted grey vertical lines represent the weekly windows.
}
\label{Fig4}       % Give a unique label
%\vspace{-0.4cm}
\end{figure*}

%%%%%%%%%Fig 2
We have embedded each of these weekly networks using the well-known DeepWalk algorithm on $D =32$-dimensional space. For the details of the embedding 
technique, see Method section. 
This gives a $D$-dimensional vector $V_i^\alpha$ for each node of the networks. We use $i, j$ as node indices and Greek letters $\alpha, \beta$ as components of the vectors on a $D$ dimensional space.  
%-----define cor tensor
In weekly networks of XRP transactions, we found $N = 71$ nodes does at least one transaction every week.  We call these $N=71$ nodes, regular nodes. Each regular node of weekly networks in the embedding space is represented by a D-dimensional vector time series $V_{i}^\alpha (t)$, where
$i=1, 2, 3,...N$, $t=1,2,3,...T$ and $\alpha=1, 2, 3,...D$. We have chosen $D = 32$ for our study. Other values of $D$ give qualitatively similar results. Later, we provide the quantitative dependence on $D$ in Eq.~\ref{eqn11}.

The correlation tensor between the different components of the regular nodes is defined as
%$M(t) = \langle V_{i}^\alpha (t) V_{j}^{\beta} (t) \rangle$, where $i,j =1,2,3, ...N$; $\alpha, \beta =1,2,3,...d$; and the $\langle \cdot \rangle$ represents
%time average for a moving time window of 3 weekly $\{t-1, t, t+1\}$ networks. 

\begin{equation}
M_{ij}^{\alpha\beta}(t) =\frac{1}{2\Delta T}\sum\limits_{t^\prime=t-\Delta T}^{t+\Delta T}\frac{[V_{i}^\alpha (t^\prime) - \overline{V_{i}^\alpha}][V_{j}^\beta (t^\prime) - \overline{ V_{j}^\beta}]}{\sigma_{V_i^\alpha} \sigma_{V_j^\beta}},  
\label{eqn1}
\end{equation}
where $\sum$ is taken over $5$ weekly $\{t-2, t-1, t, t+1, t+2\}$ networks with  $\Delta T=2$ for our analysis. The $ \overline{V_{i}^\alpha} $ and $ \sigma_{V_i^\alpha}$ represents mean and standard deviation of $V_{i}^\alpha$ over a time window of $(2 \Delta T + 1) = 5$ weekly $\{t-2, t-1, t, t+1, t+2\}$ networks.  Note that lower the values of $\Delta T$, more noise is present in the obtained correlation tensor. We also can not take $\Delta T$ too large as we are studying the dynamical evolution of the networks. We further discuss the dependence of the correlation tensor on the time window in the SI Text 2. 

%-----

%From these weekly snapshots of the vectors, we calculate the correlation tensor  $M_{ij}^{\alpha\beta}(t) $ for regular nodes as prescribed in the Method section. 
%To separate the noise from the signal 
To understand the significance of the empirical correlation tensor, we consider the following two null hypotheses. 
The null hypothesis of the components of embedding vectors for regular nodes, $V_i^\alpha$ (randomize), are independent, uniformly distributed, random variables within $[-1,1]$. We obtained the randomized correlation tensor from  $V_i^\alpha$ (randomize). Another method to randomize the empirical correlation tenor is to remove the correlation present between the different components of the embedding vectors, by reshuffling the positions of the components of embedding vectors  $V_i^\alpha$(reshuffle). We obtain the reshuffle correlation tensor from  $V_i^\alpha$ (reshuffle). The distributions of the elements of empirical, randomize and reshuffle correlation tensor are shown in Fig.~\ref{Fig2} for the week of 2017, November 06 - November 12.
We have taken the average of the data for the distributions with $20$ different runs of the embedding algorithm. 
The distributions from randomized and reshuffled correlation tensors are qualitatively identical. This is because the auto correlation for different components of the empirical embedding vectors vanishes when the lag is more than one week. These two distributions are symmetrical around zero, with an average value of the elements close to zero. The empirical correlation differs from the randomized and reshuffle correlation tensor. It is an asymmetric distribution having an average value of the elements $ 0.017$. 
 The correlation tensor has the dimension $N \times N \times D \times D$. Since we have many elements in the correlation tensor, we uncover the crucial information by diagonalizing it using a double singular value decomposition (SVD) method.  The double SVD is a natural extension of the single SVD
 for the correlation of standard non-embedded vectors $V_i$.  Since
  indices $i$ and $\alpha$ correspond to the individual node and the embedding dimension, respectively, it is also natural to 
 carry out SVD successively with the $(i,j)$-pair and with the $(\alpha,\beta)$-pair separately as 
 discussed in the section `Data and methods'.  

%%%%%%%%%%%

%%%%%%Fig3
A double SVD of the correlation tensor gives us $N \times D$ singular values $\rho_k^\gamma$, where $i=1, 2, 3, ...N$ and $\gamma =1, 2, 3, ...D$. 
For the details, refer to the Method section. 
We compare the singular values $\rho_k^\gamma$ of the empirical correlation tensor with the singular values of randomized correlation tensor $\rho_k^\gamma$~(randomize) and reshuffled correlation tensor $\rho_k^\gamma$~(reshuffle). The comparison is shown in Fig.~\ref{Fig3} for the week of 2017, November 06 - November 12. This shows that the largest singular value lies beyond the largest singular value of the randomized correlation tensor. Also, the spectral gap $(\rho_1^1 -\rho_2^1)$ in the empirical correlation tensor is significantly large compared with its random counterpart.
%Furthermore, we note that a few more ( $5$-th to $12$-th largest) singular values are significantly larger.  

%%%%%%%%%%Fig4
The temporal variation of the largest singular values of empirical $\rho_k^\gamma (t)$ and reshuffle $\tilde{\rho}_k^\gamma (t)$ correlation tensor for different weeks is shown in Fig.~\ref{Fig4}. It is evident that the largest singular value $\rho_1^1 $ is well above the largest singular values  $\tilde{\rho}_1^1 $ of the randomized correlation tensor for all weeks. However, the second largest singular value $\rho_2^1 $ lies below the randomized counterpart.  Similar to the largest singular value $\rho_1^1 $, the spectral gap $(\rho_1^1-\rho_2^1)$ appears significantly higher than that of the randomized case.  Moreover,  we observe $\tilde{\rho}_k^\gamma (t)$ remains approximately constant with time. In a stark contrast, the largest two  empirical singular values and as well as a few more singular values (not shown) vary distinctively for different weeks. 
Note that normalizing the spectral gaps by their maximum value would make the spectral gaps of
the empirical and randomized data comparable in scale, but it would not adequately demonstrate the
difference in their magnitudes. 
%%%%%%%%

\begin{figure*}
%\centering
\includegraphics[width=0.98\textwidth]{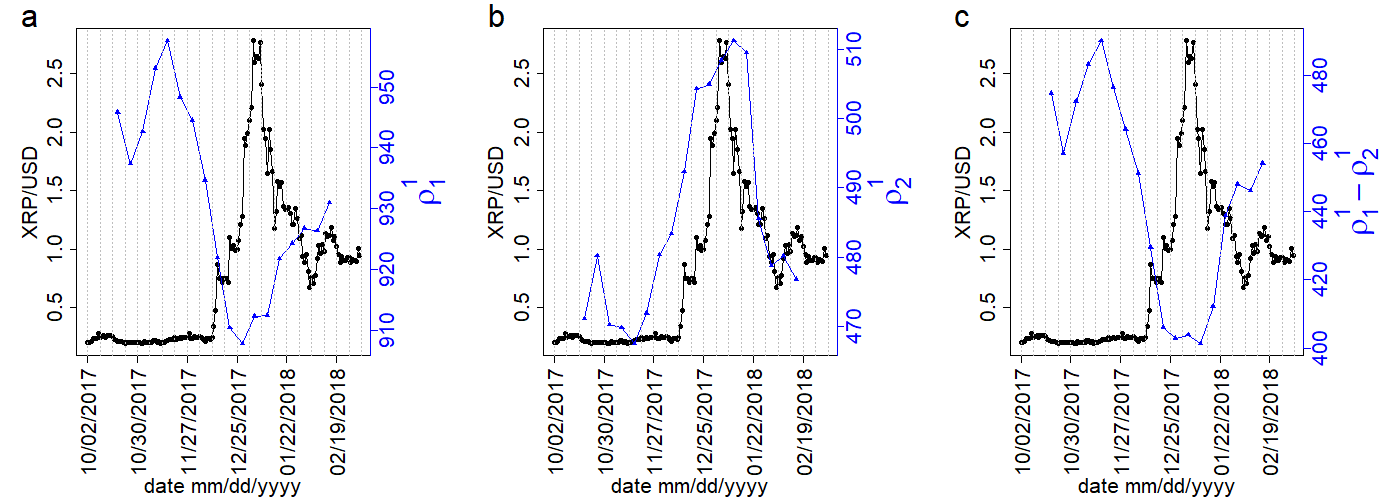}
\caption{ The comparison of daily XRP/USD price (black curves) with (a) the largest singular value $\rho_1^1$, (b) the second largest singular value  $\rho_2^1$  and (c) the spectral gap $(\rho_1^1-\rho_2^1)$ of correlation tensors for different weeks (blue curves). 
The dotted grey vertical lines represent the weekly windows.}
\label{Fig5}       % Give a unique label
%\vspace{-0.4cm}
\end{figure*}

%%%%%%Fig5
To investigate the relationship between singular values $\rho_k^\gamma$ and XRP/USD price, we compare the variation of XRP/USD daily price with the largest singular value $\rho_1^1$,  the second largest singular value $\rho_2^1$ and the spectral gap $(\rho_1^1 - \rho_2^1)$ in  Fig.~\ref{Fig5}~(a)-(c).
To quantify the dependence, we measure the correlation between the weekly XRP/USD price and the two largest singular values $\rho_1^1$, $\rho_2^1$ respectively. The weekly XRP/USD price indicates the average daily closing price of XRP/USD for the week. Let us denote the weekly XRP/USD price as $\overline{\rm XRP/USD}$.
The Pearson correlation between $\rho_1^1 (t)$ and  $\overline{\rm XRP/USD} (t+1)$ is found $r =  -0.908$ and  p-value $= 1.912 \times 10^{-7}$. The Pearson correlation between $\rho_2^1$ and $\overline{\rm XRP/USD} (t+1)$ is found $r =  0.847$ and  p-value $= 9.22 \times 10^{-6}$. Furthermore, we perform a multi-linear regression of
$\overline{\rm XRP/USD}(t+1) \sim C_0 +C_1 \rho_1^1(t) + C_2 \rho_2^1(t) $, gives $C_0 =27.91, C_1 = -0.033, C_2 = 0.008 $ and only the variable $\rho_1^1(t)$ is significant. We found $R^2=  0.8091$ and p-value $= 1.581 \times 10^{-6}$, indicating that the $80\%$ variation of $\overline{\rm XRP/USD} (t+1)$ can be explained by the largest singular value $\rho_1^1(t)$.  We also observe there are significant correlation between singular values and  weekly XRP/USD price with two weeks lead $\overline{\rm XRP/USD}(t+2)$ and three weeks lead $\overline{\rm XRP/USD}(t+3)$, which is described in detail in SI Text 3. Correlation between  $\overline{\rm XRP/USD}(t+3)$ and  $\rho_1^1(t)$ is found to be $r = -0.68$ and p-value $= 0.001$. This indicates that the largest singular value $\rho_1^1$ can give an early signal for XRP/USD price. 

\begin{figure*}
%\centering
\includegraphics[width=0.98\textwidth]{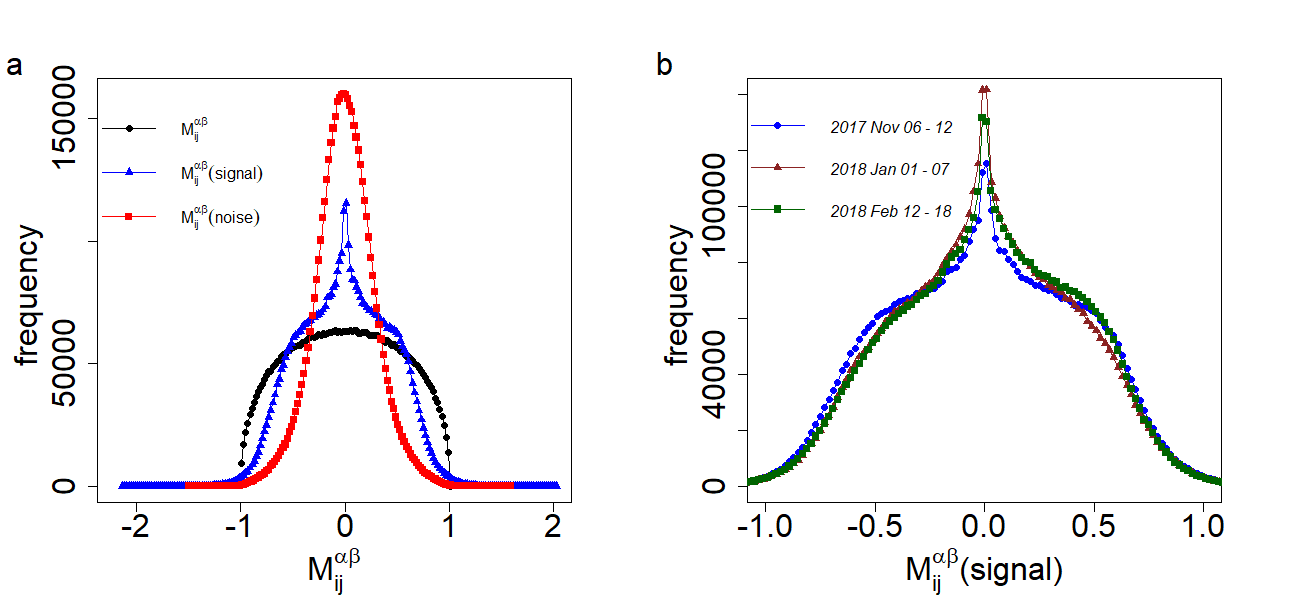}
\caption{The decomposition of the correlation tensor in the signal component and noise component. (a) Distributions for the elements of the correlation tensor, signal component and noise component for 2017, November 06 - November 12.   (b) Distributions for signal components of the correlation tensors before the bubble period ( 2017, November 06 - November 12), during the bubble period (2018, January 01- January 07) and after the bubble period (2018 February 12 - February 18).  The legends indicate different components of correlation tensor in (a), and different periods in (b).  
}
\label{Fig6}       % Give a unique label
%\vspace{-0.4cm}
\end{figure*}

We also observe that the minimum of $\rho_1^1(t)$ appears during the week of 2017 December 25 - December 31. The decomposition of the correlation tensor $M_{ij}^{\alpha \beta}$ into signal  $M_{ij}^{\alpha \beta} (\rm signal)$and noise component  $M_{ij}^{\alpha \beta} (\rm noise)$can explain the reason for this minima. The decomposition of correlation tensor $M_{ij}^{\alpha\beta}$ for the week of 2017, November 06 - November 12
is shown in Fig.~\ref{Fig6}~(a), which indicates that the distribution of the elements of the signal component is much wider than the distribution of the
elements of the noise component. Fig.~\ref{Fig6}~(b) shows the distributions of the elements of the signal component for three different weeks 2017, November 06 - November 12, 2018, January 01- January 07 and 2018, February 12 - February 18. This reflects the fact that the distribution for 2018, January 01- January 07 has a largest peak at zero and it is narrower than the other distributions, indicating that the dependence among the components of node vectors decreases during this time. The peakedness of each distribution can be quantified by its fourth moment, which is known as kurtosis.
%The kurtosis for the three distribution are found to be 2.400, 2.551 and 2.522 respectively. The standard deviation of each of these distributions are found 0.426, 0.400 and 0.406 respectively.  
The different moments for the elements of the signal component of correlation tensor are tabulated in Table~\ref{tab1}.
Clearly, the peakedness and spread of the distribution during 2018, January 01- January 07 are relatively higher and thinner, respectively. 
%sd 0.4262547, 0.3998464,  0.4059945
% 2.399955,  2.550758, 2.522026
As the average dependence among the components of the node vectors decreases during 2018, January 01- January 07, $\rho_1^1(t)$ has the minima during this period. 

\begin{figure*}
%\centering
\includegraphics[width=0.98\textwidth]{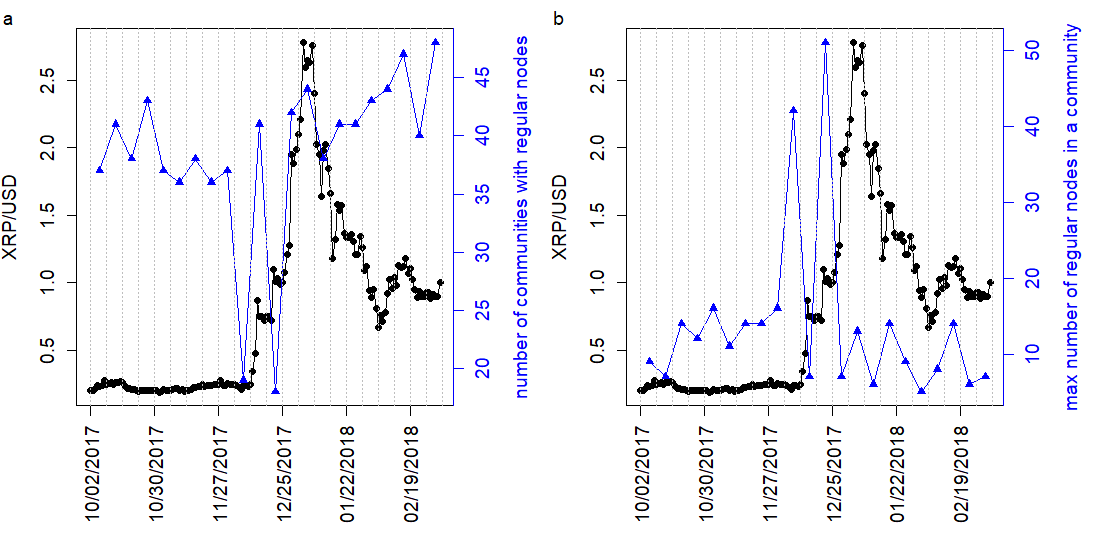}
\caption{ Evolution of regular nodes within the community structure. (a) Evolution of the number communities of the weekly XRP network that contains at least one regular node (blue curve).    (b) Evolution of the maximum number of regular nodes in a community of the weekly XRP network (blue curve).  
The black curves in both the panels represent the daily XRP/USD price. The dotted grey vertical lines represent the weekly windows.
}
\label{Fig7}       % Give a unique label
%\vspace{-0.4cm}
\end{figure*}

%%%%%%%%%%%%%%%%%%%
\begin{table}[h!]
\caption{ The mean (for the absolute values of the elements), standard deviation and kurtosis of the signal components of correlation tensors for three different weeks. The values within the brackets indicate standard deviations of the quantities. }
\centering
\begin{tabular}{|l| c| c| c|} 
 \hline
  & 2017, November 06 - November 12 & 2018, January 01- January 07 & 2018, February 12 - February 18 \\ 
 \hline
 $\langle{|M_{ij}^{\alpha\beta}(\rm signal)|}\rangle$ & $0.343 (0.006)$  &$ 0.324 (0.005)$ & $0.334 (0.005)$ \\ 
 \hline
 standard deviation & $0.419 (0.006)$ & $0.401 (0.005)$ &  $0.409 (0.005)$ \\
 \hline
 kurtosis & $2.447 (0.035) $ & $2.538 (0.032)$ & $2.500 (0.029) $ \\
 \hline
\end{tabular}
\label{tab1}
\end{table}

%%%%%%%%%%%%%%%%%%%%%%%

%%%%%% Why the correlation decreases
To understand why the average dependence among the components of node vectors decreases during the bubble period, 2018, January 01- January 07, we investigate the change in community structure of the regular nodes. We use well known Infomap algorithm~\cite{rosvall2008maps} to detect the communities in the entire weekly directed weighted networks. 
We observe the evolution of the number of communities of the XRP weekly network that contains at least one regular node in Fig.~\ref{Fig7}~(a). It is observed that the number of such communities decreases from around $40$ to around $20$ during the week 2017 December 04 - December 10 and the week 2017 December 18 - December 24. At the same duration, we observe that the maximum number of regular nodes in a community increases from around $10$ to around $50$, as shown in Fig.~\ref{Fig7}~(b). We further delve deeper into the community structure and observe the evolution of each regular node within the communities of the weekly networks. The evolution is shown in SI Fig. S3- S6. It shows that a big community of regular nodes forms during 2017, December 18 to December 24, as shown in SI Fig.~S4-~S5. This big community got fragmented in subsequent weeks. Moreover, the large community of the regular nodes remains almost intact during the non-bubble period. This distinctive change in the community structure might be related to the decrease in the average dependence among the components of node vectors or the appearance of minima for the largest singular value $\rho_1^1(t)$ during the bubble period compared with non-bubble period.  

%%%%%%%%%%%%%%%%%%%

\begin{figure}
\centering
\includegraphics[width=0.5\textwidth]{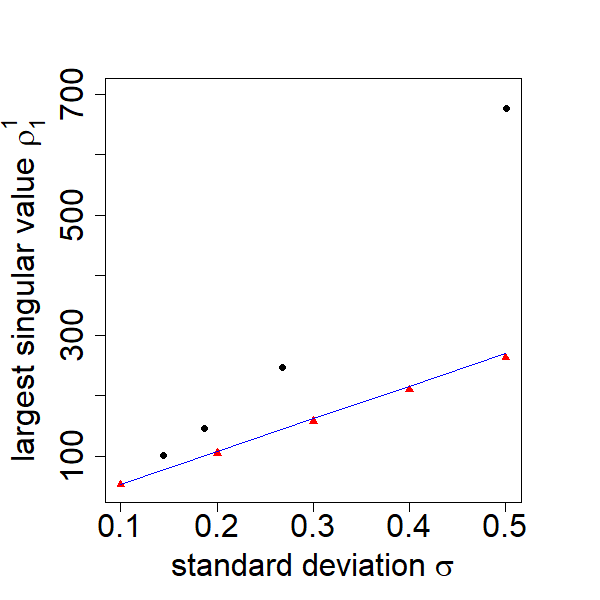}
\caption{Plot for the largest singular value $\rho_1^1$ with standard deviation of the correlation tensor elements $\sigma$. Black circles represent singular values for the randomized correlation tensors with time window $(2 \Delta T +1) = 5, 15, 30$ and $50$ (from right to left). Red triangles represent the largest singular values 
for correlation tensors, where the elements are drawn from a normal distribution with mean zero and standard deviation $\sigma$. The blue line represents $\rho_1^1 = 2 \sqrt{N} D \sigma$ which is the analytical expression for the singular values as given in Eq.~\ref{eqn11}. 
}
\label{Fig8}       % Give a unique label
%\vspace{-0.4cm}
\end{figure}

%%%%%%%%%%%%%%%%%%%%
Finally, using RMT~\cite{sengupta1999distributions, rudelson2010non, bryc2020singular, edelman2005random}, we show the relationship between the largest singular value and the standard deviation for the correlation tensor with normally distributed elements in Fig.~\ref{Fig8}. For the detail calculation, see the Method section. It also shows the deviation of the largest singular values for randomized correlation tensors calculated with different time windows. It is observed that as we increase the time window to measure the randomized correlation tensor, the largest singular value $\rho_1^1$ approaches the theoretical value $\rho_1^1 = 2 \sqrt{N} D \sigma$. 
It reflects the fact that the noise arising from the smaller time window gradually decreases as we
increase the window size.

%%%%%%%%%%%%%%%%%%%%%

%Up to three levels of \textbf{subheading} are permitted. Subheadings should not be numbered.

%\subsection*{Subsection}

%Example text under a subsection. Bulleted lists may be used where appropriate, e.g.

%\begin{itemize}
%\item First item
%\item Second item
%\end{itemize}

%\subsubsection*{Third-level section}
 
%Topical subheadings are allowed.

\section*{Conclusion}
%Traditionally RMT~\cite{sengupta1999distributions, rudelson2010non, bryc2020singular, edelman2005random, bouchaud2009financial} is extensively used to study time series data. It is used in different fields -  image analysis, genomics, epidemiology, physics, engineering,economics and finance. The aim of this method is to analyze high dimensional data to find key factors for the collective dynamics of many quantities.  For example, it is used to study the daily returns of different stocks~\cite{plerou2002random, laloux1999noise} and foreign exchange rates~\cite{chakraborty2018deviations, chakraborty2020uncovering }, monthly macroeconomic data~\cite{kichikawa2020interindustry} , or different medical data~\cite{schindler2007assessing}. Note that these are time series of macro variables that emerge due to interactions of different entities of the system. In most cases, we lack detail information of the interactions at the micro level. For example, we do not have detail information about the interaction among the individuals in stock markets.  However, in the case of XRP, we know the information for all transactions between wallets, Using this high-quality micro level data, we have extended the RMT approach for XRP transaction network. 

In this work, we have used all the direct transactions between XRP wallets in a week to construct a weekly weighted directed network. 
Using the deep walk method, we have embedded weekly snapshots of the network in vector space. Deep walk encodes community information into nodal vectors using truncated random walks on the network. Once can explore other methods of embedding, such as node2vec, to study other regularities of the network. From the weekly snapshots of the regular nodal vector components, we calculate the correlation tensor at different time periods. We have used a double SVD to remove redundant information from the correlation tensor. The significance of our result is shown by comparing it with the results of a randomized counterpart. The evolution of the largest  singular value $\rho_1^1$ shows distinctive behavior and it correlates significantly with the XRP/USD price. The decomposition of the correlation tensor into signal and noise components shows that correlation in the signal component decreases significantly during the bubble period. We explain this decrease in the correlation during the bubble period by the evolution of community structure that shows disruptive behavior during the bubble period. The disruptive behavior of community structure has also been observed in the foreign exchange market during economic downturns~\cite{chakraborty2020uncovering}.  

In summary, we have developed a method of correlation tensor spectra from XRP transactions. 
The eigenvalue decomposition of cross correlation matrix is extensively used to analyze time series data, such as daily stock price, foreign exchange rate, etc. 
 However interactions between the agents in these systems have not been considered. On the other hand,
we construct the correlation tensor from the network structure of XRP transactions. 
Then the double singular value decomposition of the correlation tensor reveals a connection between the network structure and XRP price. The method also provides crucial insights into the bubble period of the XRP price.
While the main focus of this study is to demonstrate the feasibility of the method, further comprehensive
analysis on other bubble and non-bubble periods will be studied in the future. Overall, this method has
the potential to contribute to a better understanding and detection of bubbles in financial markets.
Moreover, this method is very general and can be used to analyze the transactions of other assets.

\section*{Data and methods}

\label{method}
\subsection*{Data description}
We collected all the direct transactions between different XRP wallets from October 2, 2017 to March 4, 2018, which were recorded as ledger data using the Ripple Transaction Protocol.
We grouped these data into $T = 22$ weekly XRP networks where wallets are the nodes and a direct transaction from a
source wallet to a destination wallet forms a directed link between them. 
The link weight between a pair of wallets is determined by the sum of XRP amounts for all transactions between them in a given week.
%The total transaction volume between the wallets represents link-weight. 
See~\cite{ikeda2022characterization} for the structural properties of the XRP transaction network. 

\subsection*{Network embedding}
We embedded weighted directed weekly networks using well-known node2vec method~\cite{grover2016node2vec} with parameters $p = q = 1$, which 
is a special case of the node2vec method and represents the DeepWalk method.  Based on a natural language model, these embedding methods capture structural regularities in the network. Particularly, the DeepWalk method encodes the community structure in the vector representation of the nodes.  It uses a truncated random walks to extract the neighborhood information of nodes by generating a sequence of nodes $S = \{V_1, V_2, V_3, ... V_S\}$ that
is equivalent to a sentence in natural language. Furthermore, it applies SkipGram algorithm~\cite{mikolov2013efficient} to map each node $V_j$ to its vector representation $ \Phi(V_j) \in \mathcal{R}^D$ by maximizing the co-occurrence probability of its neighbours in the random walk.

%In weekly networks of XRP transactions, we found $N = 71$ nodes does at least one transaction every week.  We call these $N=71$ nodes regular nodes. Each of the regular nodes of weekly networks in the embedding space is represented by a D-dimensional vector time series $V_{i}^\alpha (t)$, where
%$i=1,2,...N$, $t=1,2,3,...T$ and $\alpha=1,2,3,...D$. We have chosen $D = 32$ for our study. Other values of $D$ give qualitatively similar results.

%The correlation tensor between the different components of the regular nodes is defined as
%$M(t) = \langle V_{i}^\alpha (t) V_{j}^{\beta} (t) \rangle$, where $i,j =1,2,3, ...N$; $\alpha, \beta =1,2,3,...d$; and the $\langle \cdot \rangle$ represents
%time average for a moving time window of 3 weekly $\{t-1, t, t+1\}$ networks. 

%\begin{equation}
%M_{ij}^{\alpha\beta}(t) =\frac{1}{2\Delta T}\sum\limits_{t^\prime=t-\Delta T}^{t+\Delta T}\frac{[V_{i}^\alpha (t^\prime) - \overline{V_{i}^\alpha}][V_{j}^\beta (t^\prime) - \overline{ V_{j}^\beta}]}{\sigma_{V_i^\alpha} \sigma_{V_j^\beta}},  
%\label{eqn1}
%\end{equation}

%where $\sum$ is taken over $\Delta T=2$ indicating $5$ weekly $\{t-1, t-3, t, t+1, t+2\}$ networks for our analysis. The $ \overline{V_{i}^\alpha} $ and $ \sigma_{V_i^\alpha}$ represents mean and standard deviation of $V_{i}^\alpha$ over a time window of $5$ weekly $\{t-2, t-1, t, t+1, t+2\}$ networks.  Note that lower the values of $\Delta T$, more noise is present in the obtained correlation tensor. We also can not take $\Delta T$ too large as we are studying the dynamical evolution of the networks. 

\subsection*{Double singular value decomposition}
To find the spectrum of the correlation tensor, we perform a double SVD in the following way:

We conduct the diagonalization of $M_{ij}^{\alpha\beta}$ in terms of $(ij)$-index and $(\alpha\beta)$-index successively by the bi-unitary transformation or equivalently SVD. 

The first step is
\begin{equation}
M_{ij}^{\alpha\beta} = \sum\limits_{k=1}^N L_{ik}\sigma_k^{\alpha\beta} R_{kj},
\label{eqn2}
\end{equation}
and the second step is 
\begin{equation}
\sigma_k^{\alpha\beta} = \sum\limits_{\gamma=1}^D \mathcal{L}^{\alpha\gamma} \rho_k^\gamma \mathcal{R}^{\gamma\beta}.
\label{eqn3}
\end{equation}
Then, altogether we have
\begin{equation}
M_{ij}^{\alpha\beta} = \sum\limits_{k=1}^N \sum\limits_{\gamma=1}^D \rho_k^\gamma (L_{ik} R_{kj}) (\mathcal{L}^{\alpha\gamma} \mathcal{R}^{\gamma\beta}). 
\label{eqn4}
\end{equation}
Here $ \rho_k^\gamma$ is the $N \times D$ generalized singular values. Also, note that all singular values are real and positive because the correlation tensor $M$ is real.

\subsection*{Reference correlation tensors}
We examine the properties of the empirical correlation tensor by comparing the same quantity between the original correlation tensor and reference correlation tensors, which is analogous to random matrix theory. To get the reference correlation tensors, we have used two different techniques to obtain a reshuffled correlation tensor and a randomized correlation tensor in the following way: 

\subsubsection*{Reshuffle correlation tensor} We reshuffle the components of embedding vectors $V_i^\alpha (t)$ within the time window $(2 \Delta T +1)$. Then, calculate the 
correlation tensor using Eq.~\ref{eqn1} with the reshuffled embedding vector components. 

\subsubsection*{Randomize correlation tensor}
We assign uniform random numbers between $[-1, 1]$ to the components of embedding vectors and calculate the correlation tensor using Eq.~\ref{eqn1}. 

\subsection*{Signal and noise components of the correlation tensor} 
Eq.~\ref{eqn4} can be used for dimensionality reduction to capture important relationships in the data and it can be written as 
\begin{equation}
M_{ij}^{\alpha\beta} = M_{ij}^{\alpha\beta} ({\rm signal}) +M_{ij}^{\alpha\beta} ({\rm noise}), 
\label{eqn5}
\end{equation}
where 
\begin{equation}
M_{ij}^{\alpha\beta} ({\rm signal}) = \rho_1^1 (L_{i1} R_{1j}) (\mathcal{L}^{\alpha,1} \mathcal{R}^{1,\beta}),
\label{eqn6}
\end{equation}
and
\begin{equation}
    M_{ij}^{\alpha\beta} ({\rm noise}) =  \sum\limits_{k=1}^N\sum\limits_{\gamma=1 }^d \rho_k^\gamma (L_{ik} R_{kj}) (\mathcal{L}^{\alpha\gamma} \mathcal{R}^{\gamma\beta}) -\rho_1^1 (L_{i1} R_{1j}) (\mathcal{L}^{\alpha,1} \mathcal{R}^{1,\beta}). 
\label{eqn7}
\end{equation}

The signal component is calculated using the singular values that lie above the largest singular values of the randomized correlation tensor.
In our study, we find that only $ \rho_1^1$ lies above the largest singular values of the randomized correlation tensor and it provides the dominant contributions to the correlation tensor.  

\subsection*{Singular values of the correlation tensor with normally distributed elements}
Let us first consider the case of a $p \times q$ random matrix $R$, where the elements of R are normally distributed with zero mean and standard deviation $\sigma_R$. 
The singular values $s$ of such a matrix $R$ follows the following distribution~\cite{sengupta1999distributions, rudelson2010non} 
\begin{equation}
    P(s)=\frac{1}{\pi s \sigma_R^2}\sqrt{(s_{max}^2 -s^2)(s^2-s_{min}^2)}
\end{equation}
for $s_{min} < s < s_{max}$ and zero otherwise, where

\begin{equation}
 s_{max} =\sqrt{2} \sigma_R \sqrt{\frac{(p+q)}{2} +\sqrt{pq}}   
\end{equation}
and 
\begin{equation}
 s_{min} =\sqrt{2} \sigma_R \sqrt{\frac{(p+q)}{2} -\sqrt{pq}}   
\end{equation}

When $p=q=N$ and $\sigma_R=\sigma$, $s_{max}= 2 \sigma \sqrt{N} $.  Now consider the case of the $(N \times N \times D \times D)$ random correlation tensor $M_{ran}$, where its elements are normally distributed with zero mean and standard deviation $\sigma$ . The random correlation tensor $M_{ran}$ contains $D^2$ number of $(N \times N)$ random matrix $R$. Analogous to Eq.\ref{eqn2}, SVD of the random  correlation tensor will give $\sigma_k^{\alpha\beta}(ran)$, where $\sigma_1^{\alpha\beta}(ran)$ is a $(D \times D)$ matrix, where its elements are distributed with mean $2 \sigma \sqrt{N} $.
A further SVD of  $\sigma_1^{\alpha\beta}(ran)$ corresponding to Eq.\ref{eqn3}, will give~\cite{bryc2020singular}
\begin{equation}
  \rho_1^1=2 \sigma \sqrt{N} D + O (\sqrt D). 
 \label{eqn11} 
\end{equation}

We show how the largest singular values of randomized correlation tensors deviate from the largest singular values of correlation tensors where its elements are drawn from a normal distribution with mean zero and standard deviation $\sigma$ in Fig.~\ref{Fig8}.   

%The largest singular values of a $N \times N$ square random matrices with independent entries with zero mean and unit variance 
%is given by $s_1 = 2 \sqrt{N}$ when $N \to \infty$ ~\cite{rudelson2010non}.
%A further singular value decomposition of $D \times D$ random matrix with mean $s_1 = 2 \sqrt{N}$ gives $\rho_1^1 (randomize) = 2 \sqrt{N} D + O (\sqrt D)$~\cite{bryc2020singular}. For our analysis, $N =71$, $D=32$ gives $\rho_1^1 (randomize) \sim 540 $.
\subsection*{Data availability}
We collected the data from the ripple data API at \url{https://xrpl.org/data-api.html#payment-objects}.
%\bibliography{bibl}

%\noindent LaTeX formats citations and references automatically using the bibliography records in your .bib file, which you can edit via the project menu. Use the cite command for an inline citation, e.g.  \cite{Hao:gidmaps:2014}.

%For data citations of datasets uploaded to e.g. \emph{figshare}, please use the \verb|howpublished| option in the bib entry to specify the platform and the link, as in the \verb|Hao:gidmaps:2014| example in the sample bibliography file.

\section*{Acknowledgements}
We thank the members of the Kyoto Univ.- RIKEN blockchain study group  for discussions.  
 This work is partially supported by  Ripple Impact Fund 2022-247584 (5855).

\section*{Author contributions statement}
All authors conceived and designed the study. Y.I. developed the original
 conceptualization. T.H. formulated a double singular value decomposition. A.C. analyzed the data. All authors discussed the results and contributed to the manuscript. A.C. wrote the paper. All authors reviewed the manuscript.

\newpage
	\section*{Supplementary Information}

	\renewcommand{\figurename}{SI Fig.}
	%\setcounter{figure}{0}
	%------
	 \setcounter{table}{0}
        \renewcommand{\thetable}{S\arabic{table}}%
        \setcounter{figure}{0}
        \renewcommand{\thefigure}{S\arabic{figure}}%
	%-------
\subsection*{SI Text 1: The daily XRP/USD price}
\begin{figure}[h]
\centering
\includegraphics[width=0.7\textwidth]{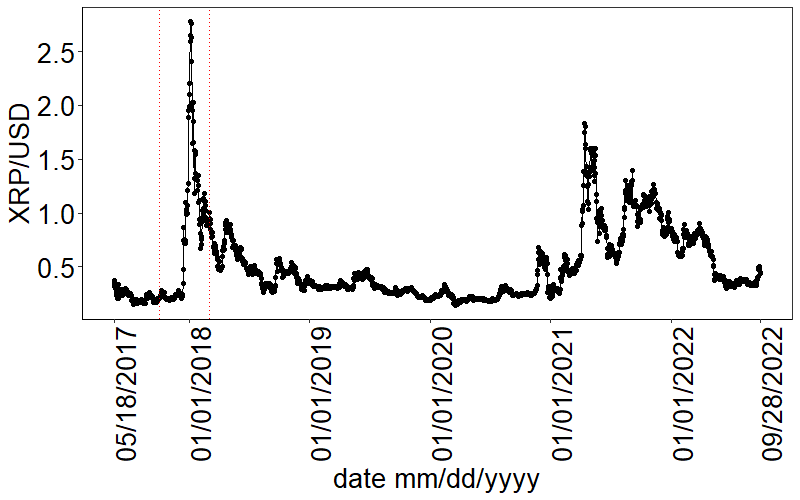}
\caption{ The daily XRP/USD close price (source: {https://www.marketwatch.com/investing/cryptocurrency/xrpusd}. The region between the red dotted lines indicates the period that we consider for our analysis.}
\label{Figs1}       % Give a unique label
%\vspace{-0.4cm}
\end{figure}

Here we show the daily XRP price between May 18 , 2017 and September 28, 2022 in Fig.~\ref{Figs1}. It can be observed that the most significant bubble formation and crash of XRP/USD rate occurred at the end of  2017 and beginning of 2018. 

\subsection*{SI Text 2: Dependence of the correlation tensor on window size}	
%The correlation tensor is defined as

\begin{figure}
\centering
\includegraphics[width=0.95\textwidth]{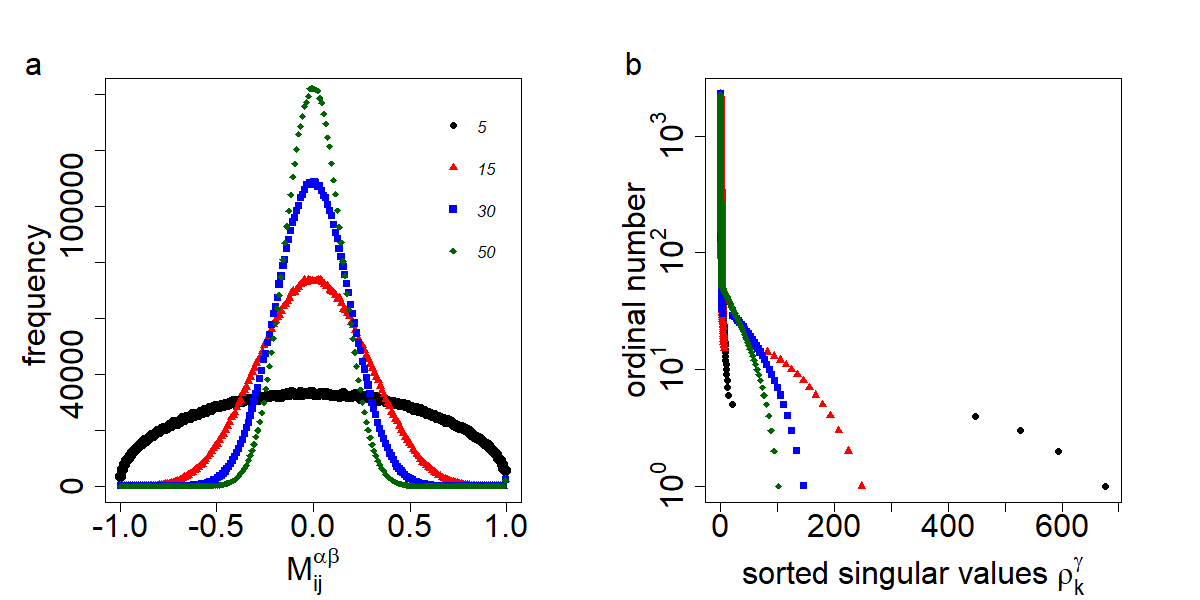}
\caption{(a) Distributions for the elements of the randomized correlation tensor calculated for different window size. The legends indicate window sizes $( 2 \Delta T+1)$. (b) Sorted singular values for the randomized correlation tensors. }
\label{Figs2}       % Give a unique label
%\vspace{-0.4cm}
\end{figure}

%$$
%M_{ij}^{\alpha\beta}(t) =\frac{1}{2\Delta T}\sum\limits_{t^\prime=t-\Delta T}^{t+\Delta T}\frac{[V_{i}^\alpha (t^\prime) - \overline{V_{i}^\alpha}][V_{j}^\beta (t^\prime) - \overline{ V_{j}^\beta}]}{\sigma_{V_i^\alpha} \sigma_{V_j^\beta}},  
%$$
%where $(\Delta T +1)$ is the window sizes and refer to the description of Eq. \ref{eqn1} in the main text for the explanation of other variables.
The distributions for the elements of the randomized correlation tensors become narrower as the time window $(2 \Delta T +1)$ increases, which is shown in Fig.~\ref{Figs2}~(a).
We also show that the largest singular value decreases as the window size increases in Fig.~\ref{Figs2}~(b). Moreover, the gap between smaller singular values and larger singular values exists for small window sizes, but vanishes as window size becomes large.

%sd(MT5)
%[1] 0.5003375
%sd(MT15)
%[1] 0.268031
%> sd(MT30)
%[1] 0.1868924
%> sd(MT50)
%[1] 0.1443476	

	\subsection*{SI Text 3: Multilinear regression model for XRP/USD}	
	We perform the following multilinear regression models for the weekly XRP/USD price and report the results in SI Table~\ref{table:1}. 
	To calculate the weekly $\overline{\rm XRP/USD}$ rate, we take $7$-day average of the daily close price of XRP/USD using the following expression
	$$\overline{\rm XRP/USD}(t) = \sum\limits_{t^\prime= t-3}^{t+3} {\rm XRP/USD}(t^\prime)$$
	
	Model 1 : $\overline{\rm XRP/USD}(t+1)$ $\sim $ intercept + the largest singular value  $\rho_1^1 (t)$ + the second largest singular value $\rho_2^1 (t)$.
	
	The model 1 explains $81\%$ of the variance (adjusted $R^2 = 0.809$) and the significant covariates is only the  largest singular value $\rho_1^1$.
	
	Model 2 : $\overline{\rm XRP/USD}(t+2)$ $\sim $ intercept + the largest singular value  $\rho_1^1 (t)$ + the second largest singular value $\rho_2^1 (t)$.
		The model 2 explains $76\%$ of the variance (adjusted $R^2 = 0.758$) and the significant covariates is only the  largest singular value $\rho_1^1$.
	
	Model 3 : $\overline{\rm XRP/USD}(t+3)$ $\sim $ intercept + the largest singular value  $\rho_1^1 (t)$ + the second largest singular value $\rho_2^1 (t)$.
	The model 3 explains only 43\% of the variance (adjusted $R^2 = 0.433$) and the largest singular value $\rho_1^1$ is weakly significant.
		
		Model 4 : $\overline{\rm XRP/USD}(t+3)$ $\sim $ intercept + the largest singular value  $\rho_1^1 (t)$.
	The model 4 explains only 47\% of the variance (adjusted $R^2 = 0.466$) and the largest singular value $\rho_1^1$ is significant.
			
\renewcommand{\arraystretch}{2}	
	\begin{table}[h!]
		\caption{ Multilinear regression table for the weekly XRP/USD price.}
		\centering
		\begin{tabular}{|l c |c| c| c|} 
			\hline
			Variables & Model 1 & Model 2 & Model 3 & Model 4 \\ 
			\hline
			Intercept & \makecell{$27.914$\\ $(14.501, 0.073)$} & \makecell{$36.502$ *\\ $(15.832, 0.036)$} & \makecell{$36.796$\\ $(25.417, 0.170)$} & \makecell{$28.489$**\\ $(7.117, 0.001)$} \\ 
			The largest singular value, $\rho_1^1$ & \makecell{ $-0.033$ ** \\ $(0.0105, 0.006)$}& \makecell{$-0.038$ **\\ $(0.011, 0.004)$}&\makecell{ $-0.035$ \\ $(0.018, 0.074)$} &\makecell{ $-0.029 $ **\\ $(0.008, 0.002)$}\\
			The second largest singular value, $\rho_2^1$ & \makecell{$0.008$\\ $(0.011, 0.446)$}&\makecell{$0.0002$\\ $(0.012, 0.986)$} & \makecell{$-0.006 $\\ $(0.018, 0.738)$}& \\
			\hline
			Observations & $18$ & $18$ & $17$ &$17$ \\
			Adjusted $R^2$ & $0.809$ & $0.758$ & $ 0.433$& $0.466 $\\
			p value of F test&  $ 1.58\times 10^{-6}$&$9.34\times 10^{-6}$ & $ 0.007$ & $0.001 $\\
			\hline
		\end{tabular}
		\begin{flushleft}
The quantities within the brackets represent standard deviations and p-values of the coefficients. We use significance codes for p value: $0~~~``***"~~~0.001~~~``**"~~~0.01~~~``*"~~~0.05~~~``~~~"~~~ 1$
%$0 \leq *** \leq 0.001 \leq ** \leq 0.01 \leq * \leq 0.05 $ 
  \end{flushleft}
		\label{table:1}
	\end{table}

\newpage	
\subsection*{SI Text 4: Evolution of regular nodes in the communities of the weekly XRP transaction networks}	
\begin{figure}
%\centering
\includegraphics[width=0.98\textwidth]{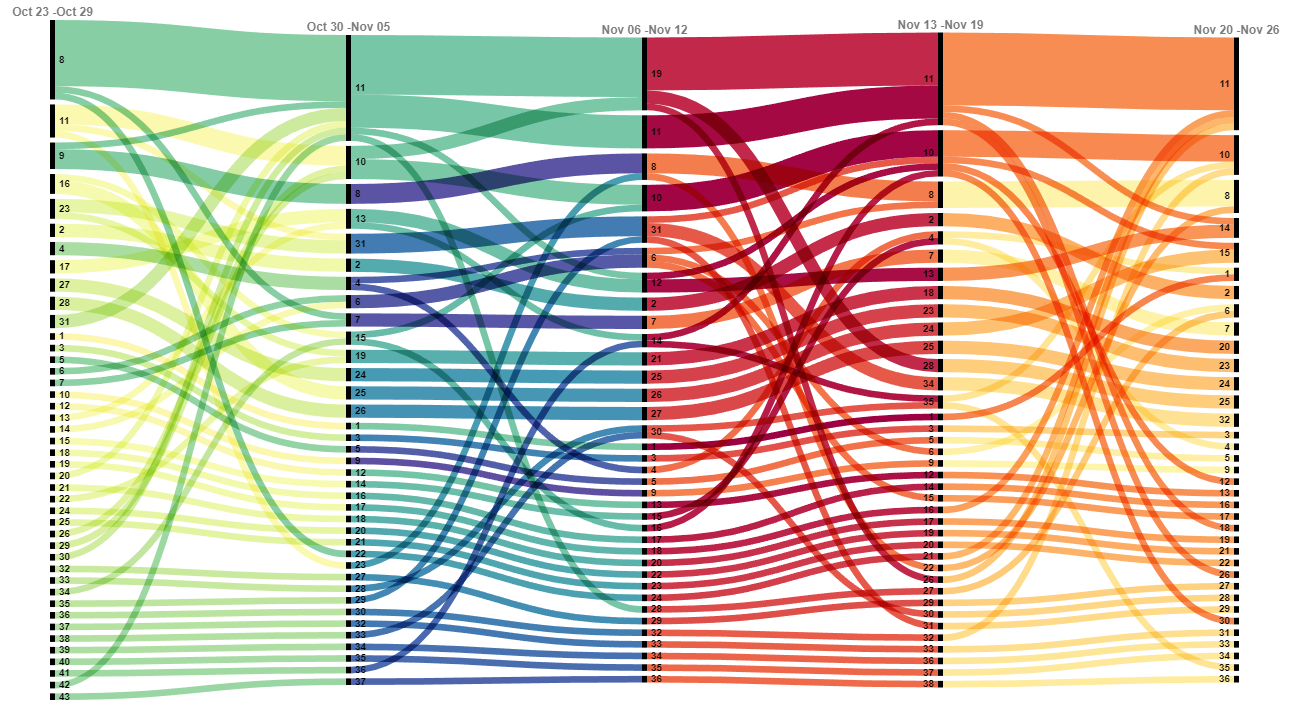}
\caption{Alluvial diagram for regular nodes showing the evolution of these nodes within the communities of $5$ consecutive weekly networks from 2017 October 23 to 2017 November 26.
Communities are arranged according to their size and the number indicates the community index. 
}
\label{Figs3}       % Give a unique label
%\vspace{-0.4cm}
\end{figure}

\begin{figure}
%\centering
\includegraphics[width=0.98\textwidth]{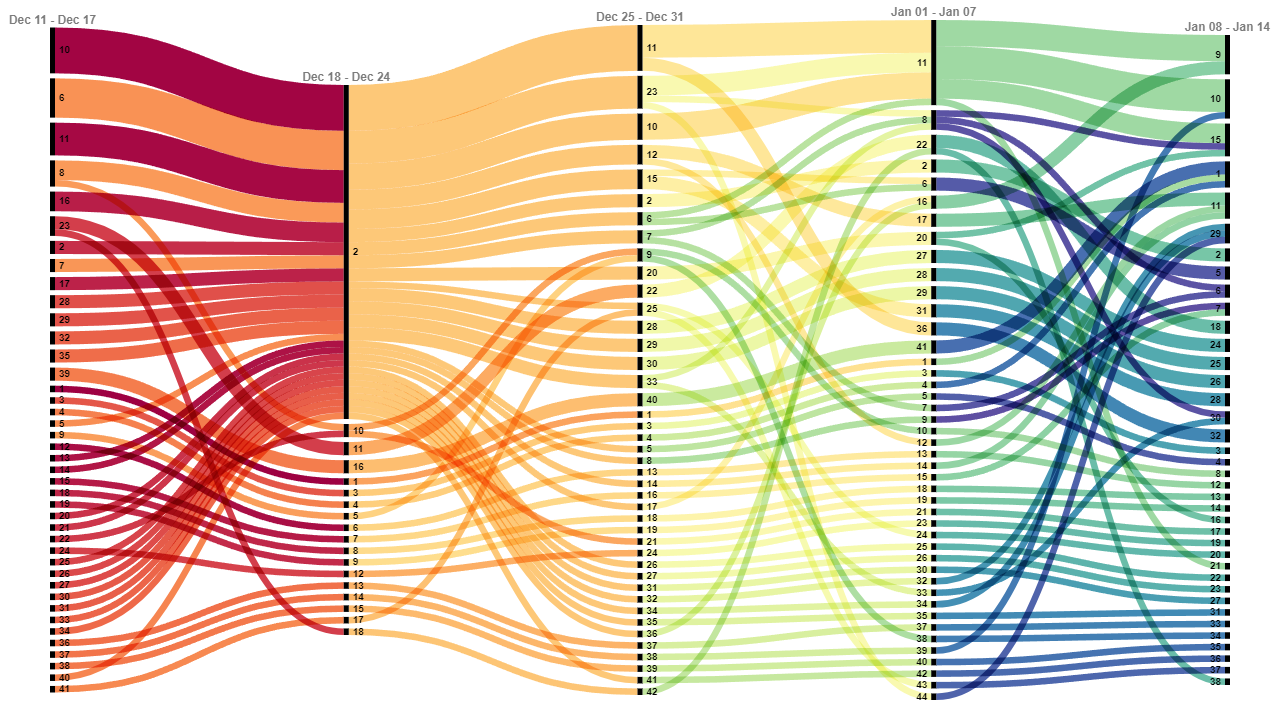}
\caption{Alluvial diagram for regular nodes showing the evolution of these nodes within the communities of $5$ consecutive weekly networks from 2017 December 11 to 2018 January 14.
Communities are arranged according to their size and the number indicates the community index. 
}
\label{Figs4}       % Give a unique label
%\vspace{-0.4cm}
\end{figure}

\begin{figure}
%\centering
\includegraphics[width=0.98\textwidth]{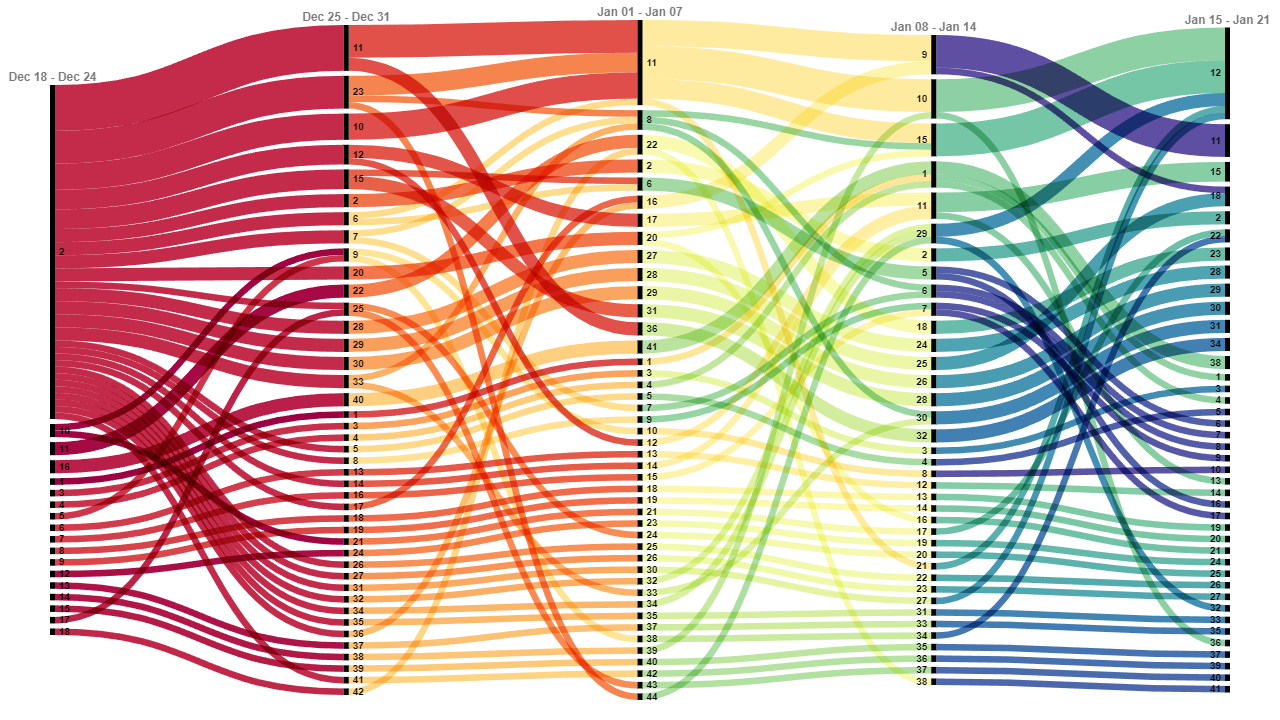}
\caption{Alluvial diagram for regular nodes showing the evolution of these nodes within the communities of $5$ consecutive weekly networks from 2017 December 18 to 2018 January 21.
Communities are arranged according to their size and the number indicates the community index. 
}
\label{Figs5}       % Give a unique label
%\vspace{-0.4cm}
\end{figure}

\begin{figure}
%\centering
\includegraphics[width=0.98\textwidth]{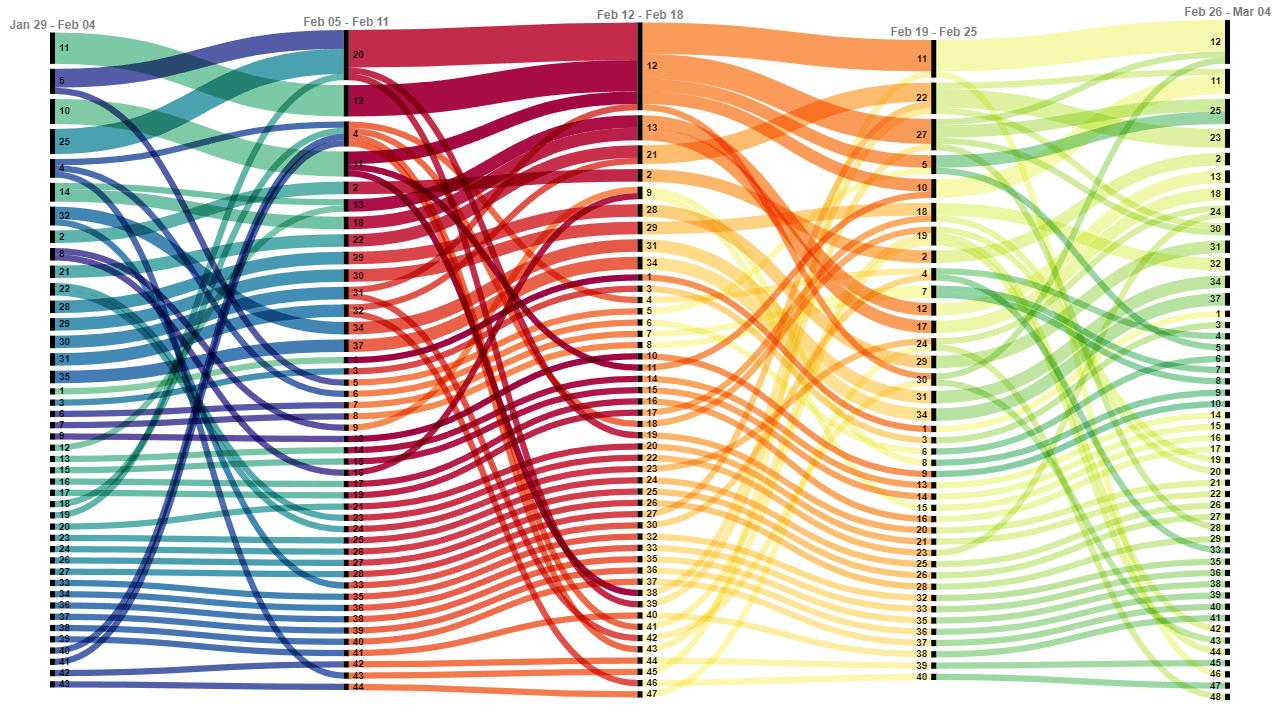}
\caption{Alluvial diagram for regular nodes showing the evolution of these nodes within the communities of $5$ consecutive weekly networks from 2018 January 29 to 2018 March 04.
Communities are arranged according to their size and the number indicates the community index. 
}
\label{Figs6}       % Give a unique label
%\vspace{-0.4cm}
\end{figure}

We use the infomap algorithm to uncover the communities for the weekly networks of XRP transactions. The evolution for the regular nodes of XRP transaction networks within the communities is shown in Fig.~\ref{Figs3}-~\ref{Figs6}.

\end{document}